# Analyzing User Perceptions of Large Language Models (LLMs) on Reddit: Sentiment and Topic Modeling of ChatGPT and DeepSeek Discussions

Krishnaveni Katta
Florida State University
kk24bh@fsu.edu

## Abstract

While there is an increased discourse on large language models (LLMs) like ChatGPT and DeepSeek, there is no comprehensive understanding of how users of online platforms, like Reddit, perceive these models. This is an important omission because public opinion can influence AI development, trust, and future policy. This study aims at analyzing Reddit discussions about ChatGPT and DeepSeek using sentiment and topic modeling to advance the understanding of user attitudes. Some of the significant topics such as trust in AI, user expectations, potential uses of the tools, reservations about AI biases, and ethical implications of their use are explored in this study. By examining these concerns, the study provides a sense of how public sentiment might shape the direction of AI development going forward. The report also mentions whether users have faith in the technology and what they see as its future. A word frequency approach is used to identify broad topics and sentiment trends. Also, topic modeling through the Latent Dirichlet Allocation (LDA) method identifies top topics in users' language, for example, potential benefits of LLMs, their technological applications, and their overall social ramifications. The study aims to inform developers and policymakers by making it easier to see how users comprehend and experience these game-changing technologies.

## 1. Introduction:

The development of Large Language Models (LLMs), such as ChatGPT and DeepSeek, has significantly transformed the artificial intelligence paradigm to introduce new innovative uses across industries from customer support to content generation. Their growing popularity, despite that, implies much remains to be understood regarding the manner in which they are perceived by their users, especially on platforms like Reddit. Public discussion of these models—particularly in off-the-record, user-generated platforms such as Reddit—is an educational means of tracking public sentiment and the general atmosphere towards these new technologies. Public sentiment plays a significant role in shaping not only future AI innovation, but also the regulatory policy that will steer it, faith in these models, and social implications of them. This study tries to fill this gap by examining Reddit discussions about LLMs with special focus on understanding what people think, believe, are concerned about, and expect. Specifically, the study will ask about common opinions on using models like ChatGPT and DeepSeek, the fear of these models being biased, and the effects of them going mainstream on a large scale.

As DeepSeek and ChatGPT take center stage on many online forums, sentiment capture and analysis of users can give useful feedback on how these models are being viewed and what are the determinants of user opinion. Such feedback is especially relevant in view of the constant stream of concerns being raised about issues like data privacy, bias, and misuse of AI.  This study examines the nuance of the views of Reddit users towards Local LLMs,interplay between the emergence of high-level AI systems and individuals' embrace of, and faith in, the systems.Sentiment analysis—a fundamental method in natural language processing (NLP)—is a great approach to solicit and measure opinions from user-generated text in this scenario. Sentiment analysis can help determine the emotional tone of the conversation on social media, classifying them as positive, negative, or neutral sentiment.

Additionally, this project paves the way for further research into sentiment analysis and AI. Future research can utilize the results of this study as a benchmark in order to come up with better sentiment analysis models and tools specific to domains such as AI, where language may be extremely technical, subtle, and context-based. By comparing both the performance of specialized sentiment analysis models and general machine learning methods, I will come to know where I need to enhance both the models as well as data on which models are trained. This research also has practical significance for organizations as well as business organizations that implement sentiment analysis for their public-based policies. With a better understanding of the public perception regarding LLMs, these stakeholders can address public concerns, build confidence in AI, and facilitate the ethical creation of future AI models. In addition to this, because AI technologies will continue to develop at a high speed, sentiment analysis's timeliness in observing and shaping citizens' opinions about them will remain increasingly significant. Through quantification of whether internet users hold positive or negative sentiments towards emerging AI technologies, scientists can make the development of the models converge to people's values and what they require.

**Research Questions:**

The primary research questions driving this study are:

- **RQ1:** What is the public sentiment of Reddit users towards the use of LLMs like ChatGPT and DeepSeek?
- **RQ2:** What are emerging topics regarding LLM?
- **RQ3:** How does the frequency of Reddit comments change over time across different quarters, and what are the patterns in the number of comments posted in relation to these time periods?

## 2. Literature Review:

In the past decade, Large Language Models (LLMs) took the natural language processing (NLP) world by storm and diversified their applications in abundance. As GPT-3, GPT-4, BERT, and the like broke records to new standards, LLMs have displayed remarkable performance in a wide array of NLP tasks from text generation to machine translation, summarization, question answering, and more (Brown et al., 2020). LLMs utilize substantial corpora of data and intricate structures to accomplish human-level performance on specific tasks and are consequently a fundamental component of contemporary AI research. Local LLMs' higher popularity can be witnessed online, where discussion is more likely to proceed in the technological features of models such as DeepSeek and Local LLaMA.

Emotional interpretation of internet conversation on Reddit provides us with a sense of the attitude of the users towards such new technology, with a dash of hope and fear of LLMs (Zhang, 2024). Local LLM like DeepSeek was developed for making advanced AI technology accessible to a greater number of people.

The users are also concerned that AI models cannot generate content in a responsible manner, especially with the risk of disseminating misinformation or causing harm unintentionally (Trandabat et al., D.2023,Rodrigues et al.,2020) A number of these conversations discuss the unaccountability of AI-generated content, and how users wonder whether someone should be held responsible when an AI system generates controversial or harmful content.

**4. Methodology:**

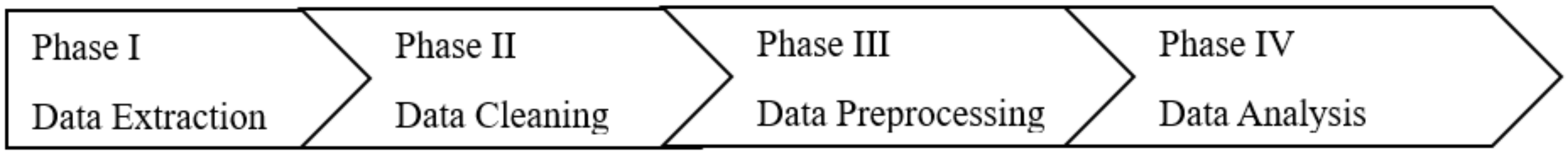

*Fig 1. Research Phases*

**Data Collection :** I tried to gather user feedback in this research from Reddit discussions that talked about Local Large Language Models (LLMs), e.g., ChatGPT, DeepSeek, and other future Local LLMs. I selected Reddit as the source of data because it is popular among AI users, researchers, and the general public, so it is a suitable platform for comment, criticism, and discussion of the developing world of AI technologies. To collect data, I used the AsyncPRAW (Asynchronous Python Reddit API Wrapper) library that provides a way to scrape comments from various Reddit threads in an efficient and asynchronous manner. I compiled a list of URLs of pertinent Reddit submissions so that the threads I chose were varied discussions about Local LLMs, their pros, cons, ethical concerns, and actual applications in the real world.

**Data Preparation:** The Reddit comments gathered were preprocessed in a number of ways to ready them for analysis before the use of sentiment analysis models. Text normalization consisted of standardizing the text format, stripping off punctuation, and making all the text lowercase in order to be consistent in the analysis. With tokenization, the comments were broken down into separate words (tokens) so that the models could learn word frequency and co-occurrence in the comments. Cutting down to regularize word variation and improve model precision entailed cutting down on words to root or base word form (e.g., "running" to "run").

**Sentiment Analysis Models:** For analyzing the sentiment of the comments that I had gathered, I utilized a variety of sentiment analysis models that each embodied different approaches in machine learning and natural language processing. They were chosen to see whether they could hold up to being utilized for analyzing the sentiment conveyed by the users on Reddit, which was likely to have colloquial language, sarcasm, and AI-specific technical terms. The models that I utilized were: Logistic Regression: A machine learning classification algorithm that is most commonly applied in binary classification problems. It was applied in this project to classify comments into positive, neutral, or negative sentiment based on whether certain words are present and their respective weights. Naive Bayes: A Bayes' theorem-based probabilistic classifier that is aptly suited for text classification problems by making an assumption of features being independent. It was used to classify sentiment on the basis of the frequency of occurrence of words and their probability in each sentiment class. Support Vector Machines (SVM): A powerful supervised learning technique for classification. SVM was utilized because SVM can manage high-dimensional data and decide boundary lines between sentiment classes. SVM with a linear kernel was utilized in our study for sentiment classification based on comment features. Random Forest: An ensemble learning model that combines several decision trees to predict outcomes. Random Forest was utilized for sentiment prediction through the aggregation of numerous decision tree predictions, enabling the sentiment labeling to be both more precise and robust.

VADER Sentiment: VADER is a lexicon-based sentiment analysis tool that has a particular talent for social media text. VADER is highly suited to deal with informal and emotive text on Reddit forums and thus is a perfect tool to be utilized in this research. TextBlob Sentiment: TextBlob is an easy-to-use, simple Python NLP library. TextBlob utilizes a rule-based system of sentiment analysis and was implemented to offer a baseline sentiment model with simplicity and ease of use in mind. Each of the models was run on the Reddit comments dataset, and the result was a sentiment label for every comment: positive, negative, or neutral. For the machine learning models (Random Forest, SVM, Naive Bayes, and Logistic Regression), text representation was done using bag-of-words, where every comment was tokenized and represented as a vector of word frequencies.

**5. Results:**

Here, I examine the 20 most common words in Reddit comments about large language models (LLMs) like ChatGPT and DeepSeek, from a total of 10,733 entries. The analysis gives us an idea of the users' attitudes and perceptions about these models.

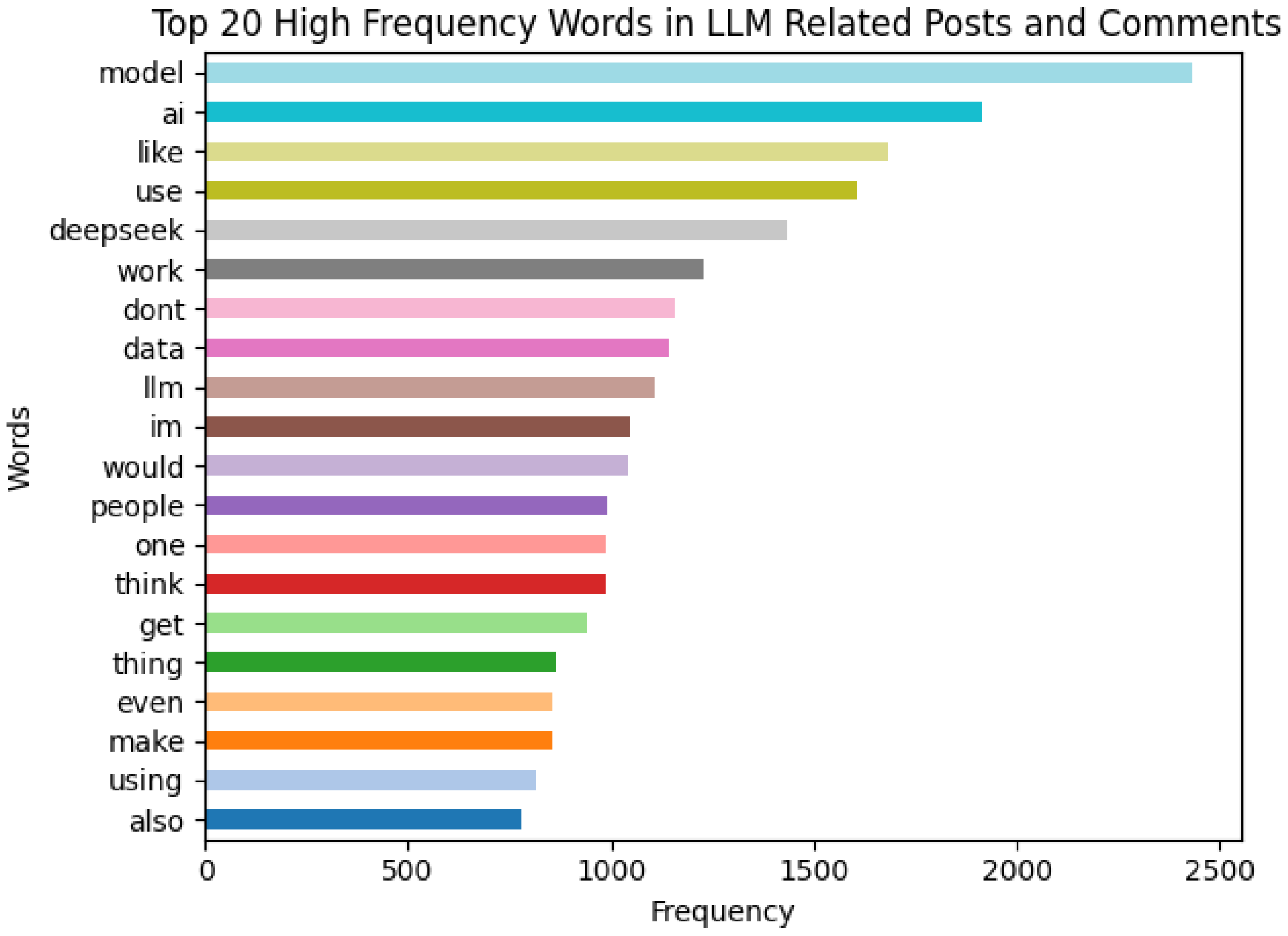

*Fig 2. illustrates the Top 20 most frequent words in the dataset.*

Words like "deepseek", "Chatgpt" and "llm" point to explicit mentions of the models in question 1. Furthermore, adjectives like "ai," "people," "data," and "make" indicate a priority on the operational uses and the real-world application these models do for people's work and activities. Self-examination is also present, as words like "would," "think," and "im" emphasize users' considerations regarding how these technologies function. In contrast, terms like "dont" and "get" indicate frustration or uncertainty, perhaps indicating issues with comprehension or using these models. Terms like "code," "model," and "using" indicate topics of the technical nature and utilization of these systems. The ubiquitous use of words like "like" and "ai" indicates that most users are interested and optimistic about LLMs, though with some apprehension. The VADER SentimentIntensityAnalyzer is used for sentiment analysis. TextBlob is considered to compare. The sentiment distribution of the 10,733 entries is shown in Fig 3; it can be noticed that 54% of the comments express positive sentiments, whereas 25.8% are negative, and the remaining 20.2% are neutral.

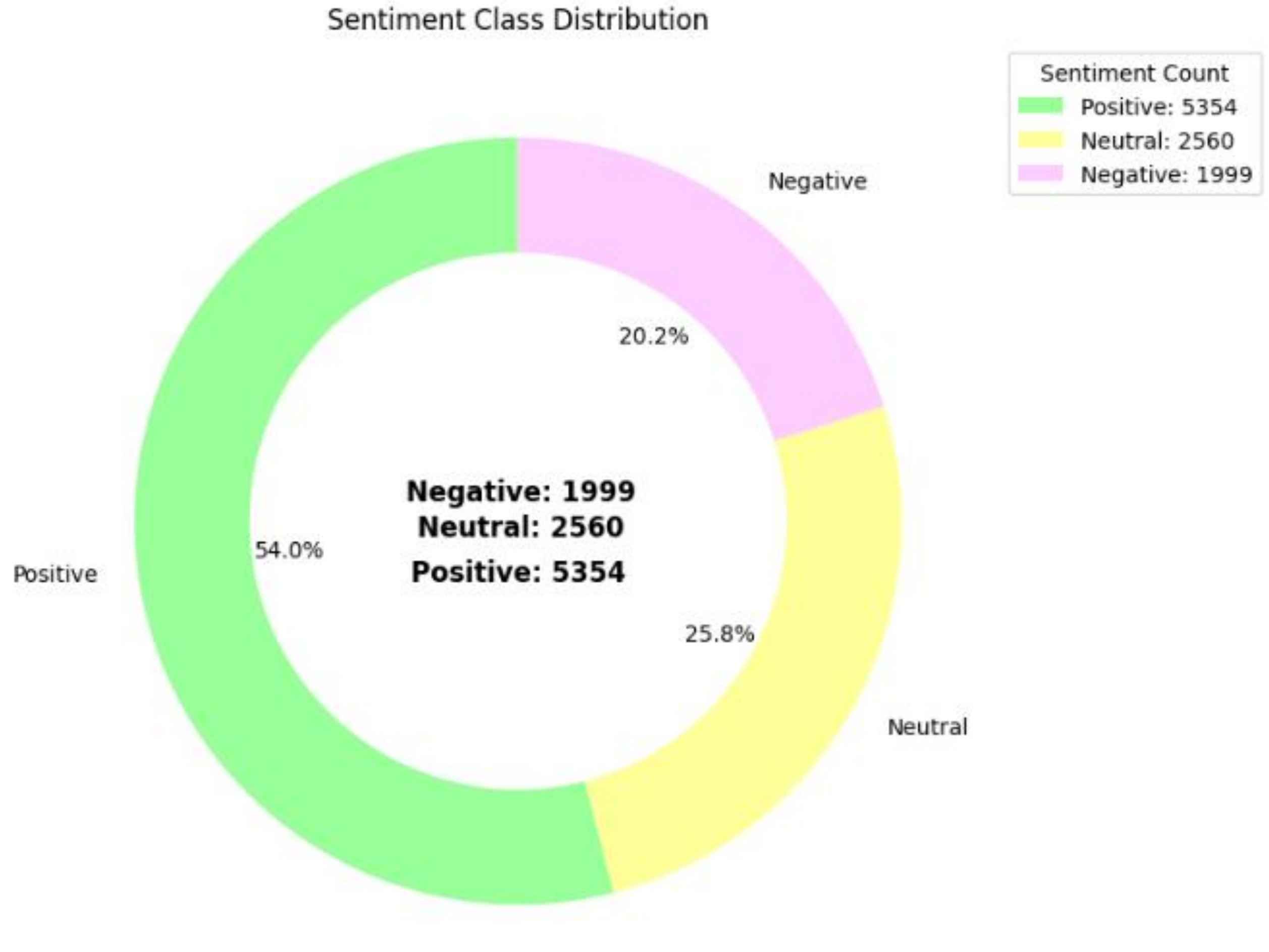

***Fig 3.*** *showing the distribution of positive, negative, and neutral sentiments towards LLM.*

This section reveals the emerging topics and themes that were identified through topic modeling. It attempts to address research question 2: What are emerging topics regarding LLM? This study combines qualitative and quantitative content analysis in an endeavor to identify and reveal latent themes and topics of public discourse, which is supposed to have immense research potential in the field of social media. As a widely used quantitative method for topic modeling, the LDA model assists in determining the optimal number of topics for classification as shown in Table 1. Nevertheless, the most perplexed is not always the best model. Models overfit if the number of topics is too high, and the result is too many and non-convergent topics. Too many topics can create excessive redundancy, which leads to low distinctiveness and uniqueness among topics. Therefore, most studies employ human judges to identify the ideal number of topics. It also has some criteria: (1) topic and word coherence being high; (2) topic quality being non-repetition, non-conflict, and addressing basic content. It assesses this topic classification and high-frequency words of every topic when the topic number was set to 8.  As the topic seems irrelevant, there are no shared themes among the topics, and the terms are not interrelated and are not associated which can be seen with low topic coherence and quality, there exists a model limitation that terms do not develop well among coherent topics.

| S.No | Topic Name | Keywords |
|---|---|---|
| 1 | Topic 1 | deepseek, ai, china, openai, u, free, chatgpt, open, company, source |
| 2 | Topic 2 | model, data, training, llm, ai, like, train, would, think, thats |
| 3 | Topic 3 | people, dont, think, even, like, u, cost, company, time, day |
| 4 | Topic 4 | model, im, run, ram, get, would, know, gpu, running, thats |
| 5 | Topic 5 | openrouter, thank, intellectual, nvidia, yes, deepseekr1, broke, commercially, number, palestinian |
| 6 | Topic 6 | like, chinese, country, court, lol, american, ask, china, got, sound |
| 7 | Topic 7 | model, deepseek, use, better, code, good, r1, o1, using, claude |
| 8 | Topic 8 | ai, work, copyright, use, make, like, human, dont, would, thing |

*Table 1. Categorization of Reddit Topics on LLMs*

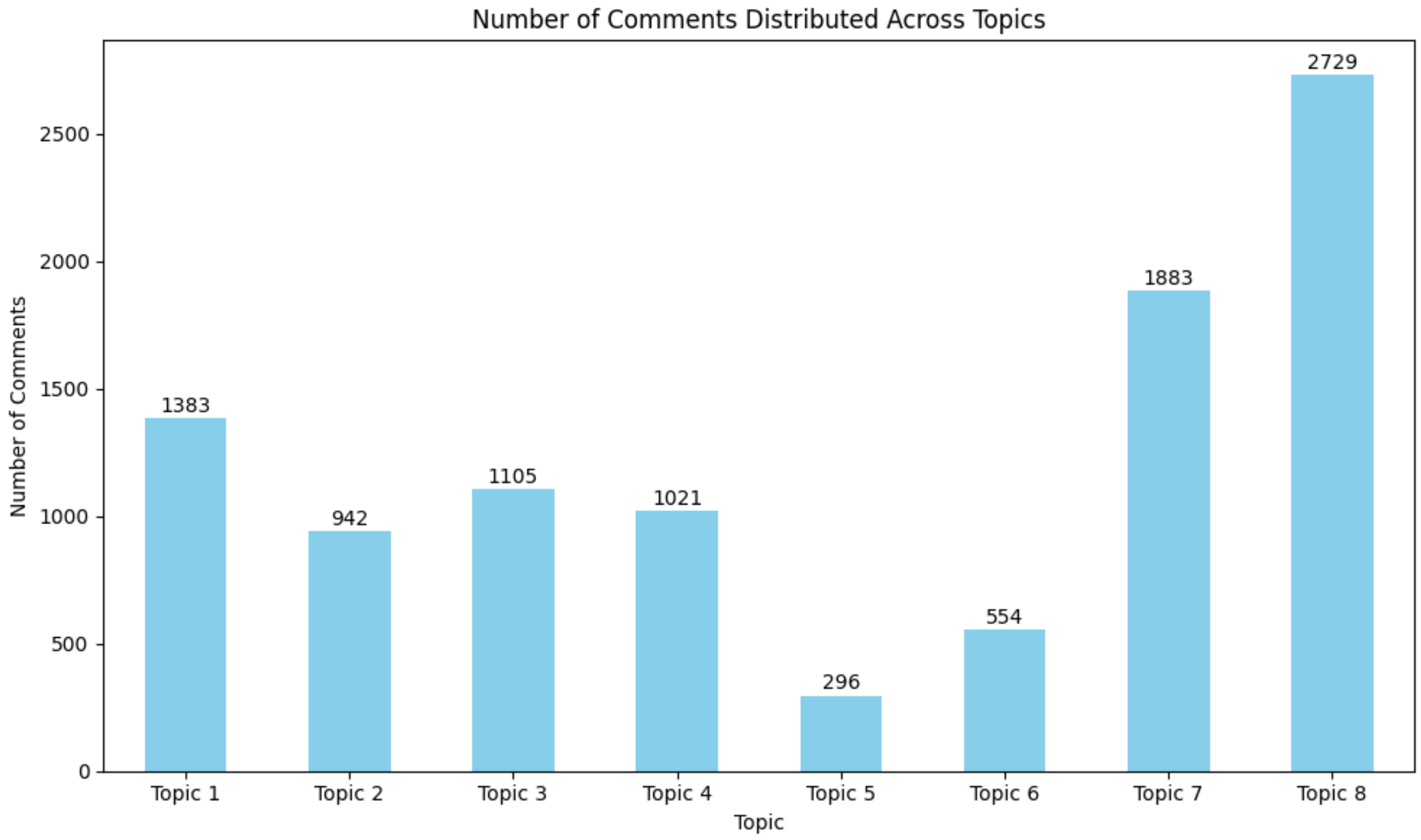

*Fig 4. showing the word frequency towards LLM.*

Therefore, this research also tried the number of topics at the point of drop in perplexity, i.e., the number of topics (7) near the point of inflection of the curve. The distribution of word frequency in relevant topics at the 7-topic setting is shown in Table 2. The most representative words within each topic have good topic quality and coherence. The most frequent words of each topic are grouped into seven topics, which further fall under three themes. The seven themes indicate the general discussion domains on LLM in the Reddit community.

| **No** | **Topic Name** | **Keywords** | **Theme** |
|---|---|---|---|
| 1 | Topic 1 | deepseek, ai, chatgpt, openai, u, data, use, source, open, company | LLM Development & Training |
| 2 | Topic 2 | ai, model, data, llm, training, use, like, would, think, new | LLM Development & Training |
| 3 | Topic 3 | china, people, u, dont, think, like, even, year, company, power | Legal and Ethical Concerns |
| 4 | Topic 4 | work, ai, make, human, know, dont, thats, im, like, doesnt | LLM Applications and user Interaction |
| 5 | Topic 5 | ram, gpu, run, nvidia, vram, cpu, gpus, context, memory, gb | LLM Applications and user Interaction |
| 6 | Topic 6 | copyright, work, book, like, law, ai, question, copyrighted, use, would | Legal and Ethical Concerns |
| 7 | Topic 7 | model, deepseek, use, better, code, im, good, run, like, r1 | LLM Development & Training |

*Table 2. Theme based Categorization of Reddit Topics on LLMs*

Seven themes were coded and categorized systematically into three themes based on qualitative content analysis. Three themes were derived from qualitative content analysis and categorized based on the topics shown in Table 2. Theme 1 of LLM Development & Training includes Topic 1 , Topic 2 and Topic 7, which highlight the technical aspects of AI and LLMs such as deepseek, OpenAI, model training, and their various applications in data handling and utilization. This theme focuses on the key technological developments and features of large language models such as ChatGPT and deepseek. Theme 2, LLM Applications and user Interaction, consists of Topic 4 and Topic 5, which speaks about how AI such as ChatGPT engages with humans, or how it affects work and human abilities. The debates in this case are on how AI might improve human decision-making, work processes, and how limited AI is when copying human capabilities. Theme 3, Legal and Ethical Concerns, includes Topic 3 and Topic 6. These concerns on the application of AI in intellectual property, more so copyright and legal concerns. This theme considers the ethical aspects of using AI in producing content and how it affects creative industries.

The baseline scores are then compared with the perplexity scores attained by the subsets sampled on the other ten occasions. A similarity test is subsequently conducted to identify the comparability illustrated in Fig 5. The horizontal axis of the heat map in this case illustrates the number of topics using LDA for topic modeling, from 2 to 10, and the vertical axis illustrates the number of subsets that we randomly sampled from the original dataset. Different squares represent the perplexity difference between the topic modeling result of 10 randomly selected subsets of the initial data set and the initial data set's topic modeling.

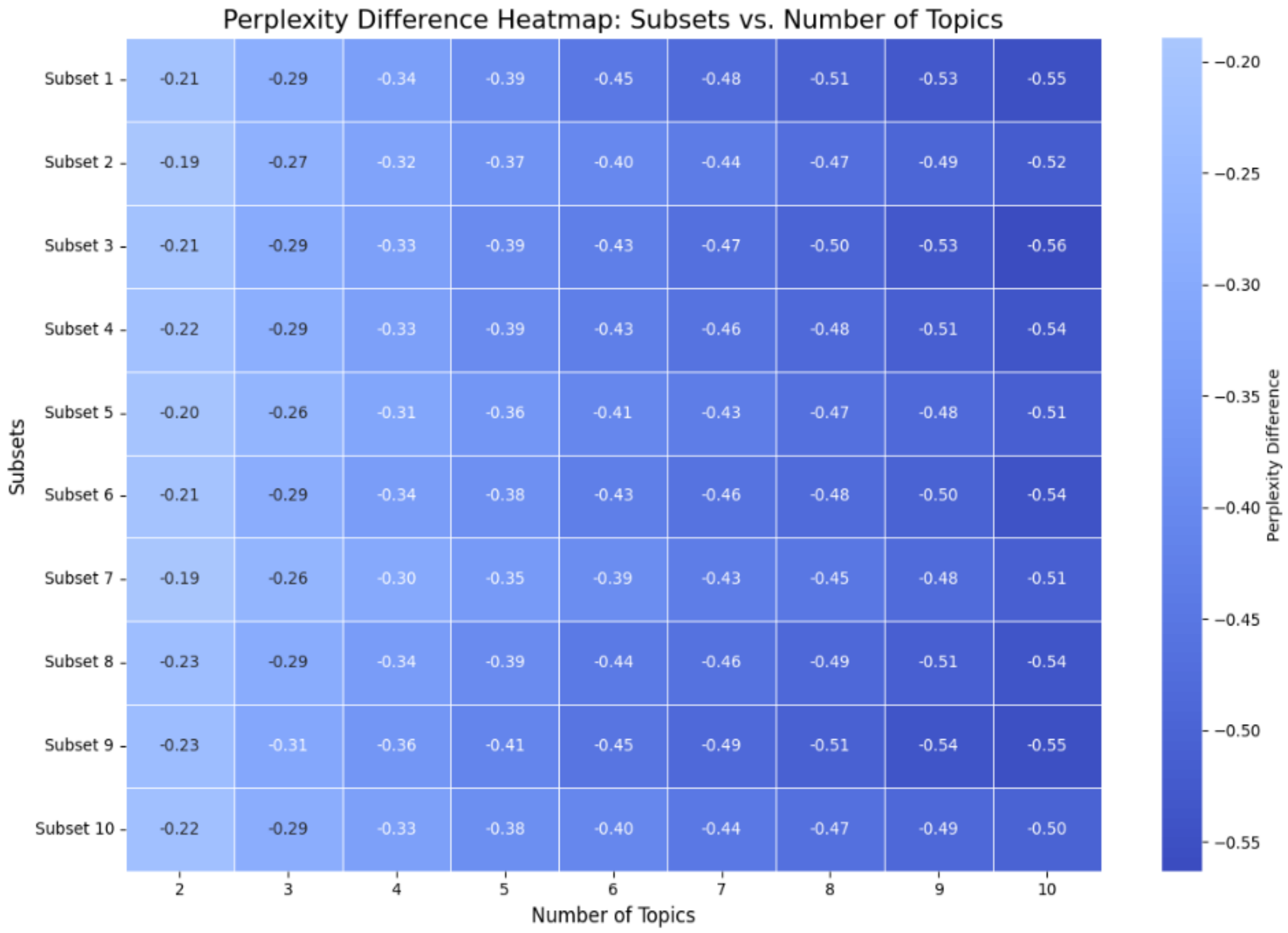

*Fig 5. Heatmaps to show the concentration of certain topics by subreddit.*

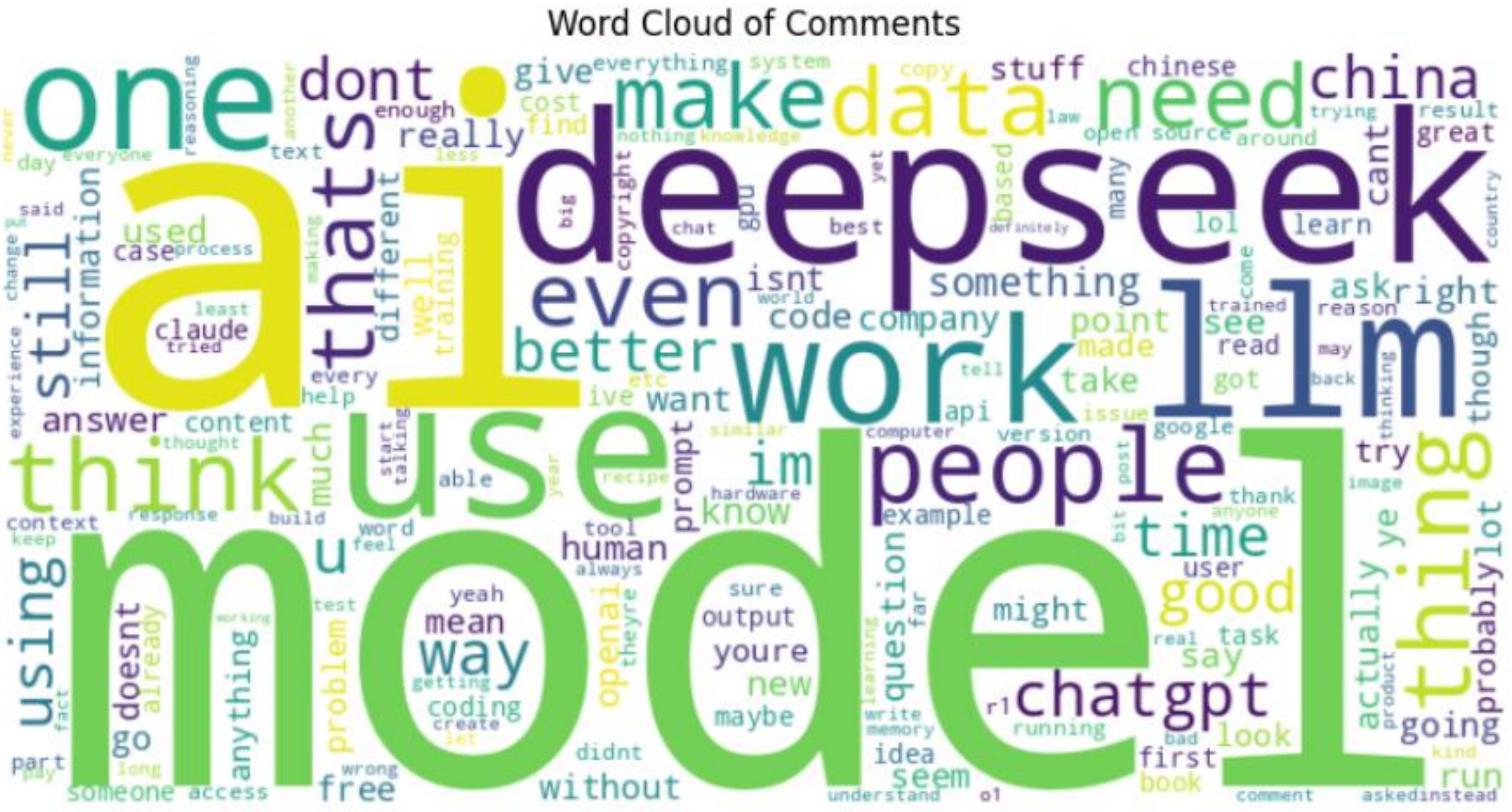

*Fig 6. Word cloud or topic distribution chart generated from topic modeling, showing the most frequently mentioned topics in Reddit discussions.*

**RQ3: How does the frequency of Reddit comments change over time across different quarters, and what are the patterns in the number of comments posted in relation to these time periods?**

Time-based trends and pattern of user activity are revealed by examining Reddit comments over time, segregated based on quarters. By examining the number of comments for each quarter, we could simply see patterns based on different intervals. In the data, all comments have dates and based on this information, the comments are segregated into quarters (i.e., Q1, Q2, Q3, and Q4). The y-axis is on a logarithmic scale to emphasize changes in the frequency of comments over time. From the bar chart, one can see that some quarters report greater activity as far as the number of comments goes, while other quarters report somewhat lower activity. Quarterly information enables us to determine if any seasonal or cyclic patterns exist that may be prompted by outside activities such as noteworthy events, popular trends, or even activities occurring within the individual subreddits.

Fig 7. illustrates the following findings: The rate at which Reddit is commented on considerably differs from one quarter to the next. There are dominant peaks in certain quarters, which indicate greater user activity. The employment of a logarithmic scale helps to normalize the imbalance in comment numbers and renders it more practical to identify relative trends throughout the time periods.

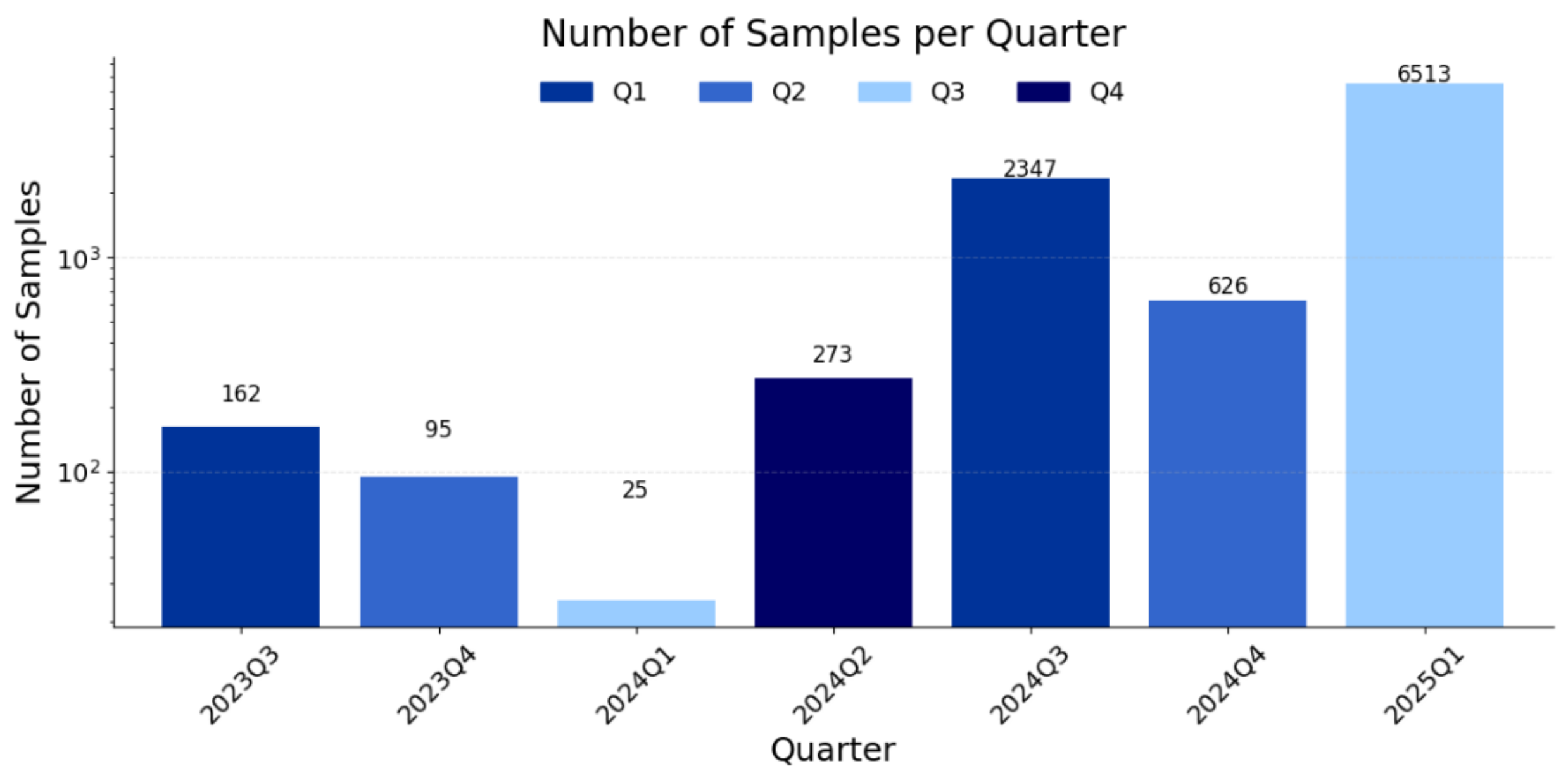

*Fig 7. Quarterly Trends in Reddit User Activity: Logarithmic Analysis of Comment Frequency*.

There are certain quarters that possess extremely sharp peaks in activity, that is: 2024Q3 (2,347 comments), the peak comment quarter. 2025Q1 (6,513 comments) also has a significantly larger overall total, reflecting a surge around this period. These Spikes reflect focused user activity on trending subjects like LLM going viral on Reddit. Another quarter similar to 2024Q1 (25 comments) and 2023Q4 (95 comments) reflect relatively low activity. The results suggest that user activity on Reddit is not evenly distributed throughout the year, with some quarters of the year having higher or lower activity than others. This could be reflective of varying levels of interest in discussion topics, external events, or other factors affecting user behavior within the network. The visualized data provides important insight into the changing nature of activity on Reddit over time and can therefore be an important tool for further examination of the causes of such variation.

Before sentiment analysis, we had tested four machine learning techniques namely Logistic Regression, Naïve Bayes, Support Vector Machines (SVM) and Random Forest and two lexicon-based techniques namely VADER and Textblob to identify the most appropriate sentiment analysis model for our dataset. I thus employed a dataset with human-labelled sentiment labels. The aim here is to contrast the outputs generated by machine learning techniques and lexicon-based techniques with human annotators' handpicked outputs. For this purpose, we used a manually annotated sentiment corpus that closely matches our specific situation, thus making a suitable choice. To compare model performance, I employed evaluation metrics including accuracy, precision, recall, and F1 score shown in Fig 8.

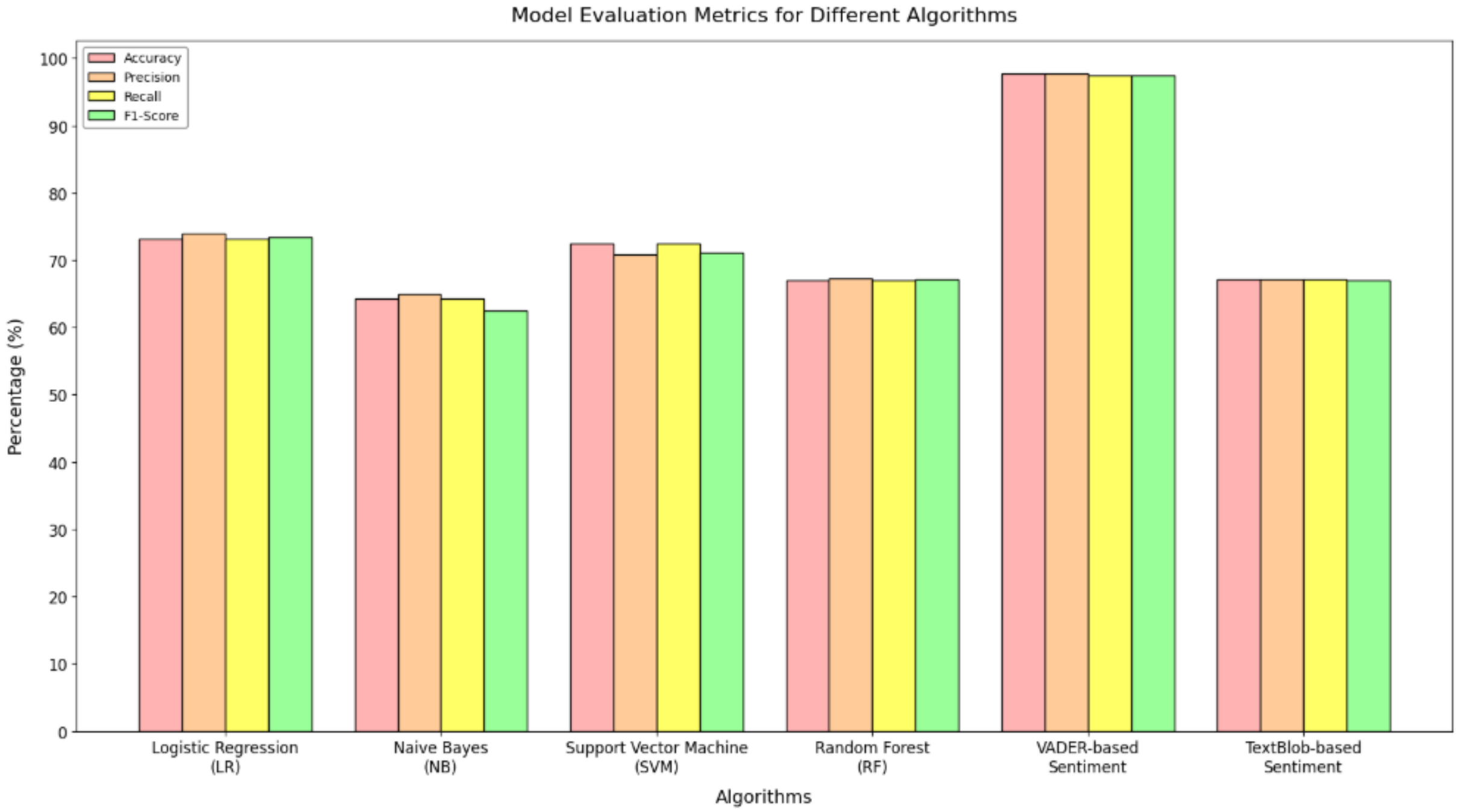

*Fig 8. Comparison of Sentiment Analysis Models: Performance Metrics Across Machine Learning and Lexicon-Based Techniques.*

VADER Sentiment performed the best across all four measurement metrics, with a highly impressive Accuracy of 97.7%, Precision of 97.7%, Recall of 97.4%, and F1 Score of 97.4%. How VADER managed to perform so well is because it is specially designed as a lexicon-based sentiment analysis tool that is specially fine-tuned for social media posts. Given the in-formal and expressive nature of Reddit comments, VADER's performance at well-classifying sentiment for short, context-dependent, and affect-laden text is particularly well-suited. Support Vector Machines (SVM) had competitive performance with Accuracy of 72.2%, Precision of 69.7%, and Recall of 72.2%. The high Recall of this model suggests that it is very good at identifying sentiment in the comments on Reddit, although its Precision was a little lower. This implies that while SVM identifies positive and negative sentiments well, it may also identify neutral comments as positive or negative. Logistic Regression performed equally well as SVM with 70.2% Accuracy and 71.1% F1 Score. The good performance of Logistic Regression shows its ability to recognize basic patterns in the text but couldn't perform better than SVM and VADER on the basis of overall accuracy to predict sentiment. Random Forest, Accuracy at 66.9%, was moderate. While the model had fairly equal Precision and Recall values, it lagged behind SVM and Logistic Regression. The marginally lower scores show that while Random Forest's ensemble approach is particularly apt for classification issues in most disciplines, it perhaps is not so fine-tuned for the nuances of language involved in posts about Local LLMs on Reddit. Naive Bayes, with an accuracy of 65.6%, was the poorest performing machine learning model. Although simple and well-suited to perform the majority of text classification tasks, it could not include the richness of sentiment present in Reddit comments about Local LLMs. The model performance was compromised by the feature independence assumption, which is not valid for most real-text data sets, especially data sets containing context-dependent phrases or slang. TextBlob Sentiment, with an Accuracy of 67.8%, was adequate but could not beat more sophisticated models like VADER and SVM. TextBlob is a simple rule-based sentiment analysis component that is based on pre-trained models and lexicons, and its performance was therefore less strong than the improved capability of VADER in handling informal social media messages.

**6. Discussion:**

The orientations of user towards ChatGPT, DeepSeek, and other LLM technologies are contrasting orientations which we are able to find by this research. These findings are consistent with previous research, which has established that how much people use AI varies according to transparency, fairness, and prevention of harmful biases. The outcomes of sentiment analysis models demonstrate the relative capacity of machine learning-driven and lexicon-driven methods for classifying the sentiment of comments. The performance excellence of VADER implies the utilization of efforts invested in the idiosyncrasies of social media content such as colloquial expressions, emoticons, shortcuts, and tags. This confirms the contention that sentiment analysis on online communities demands models attuned to these forms of communication, as demonstrated through VADER's outperforming of standard machine learning models.

These models were subsequently followed by VADER, demonstrating that machine learning models could be trained to identify sentiment but required additional fine-tuning and feature engineering to remain competitive with lexicon-based models for the detection of sentiment on the specific example of Reddit discourse. Random Forest's would be even more accurate with additional hyperparameter tuning or even more advanced features that glean sentiment signals from social text. Naive Bayes's poor performance shows the challenge of using simple probabilistic models for difficult problems. While Naive Bayes is effective and straightforward, it could not cope with the word interdependence of natural language and therefore performed relatively poorer on the Reddit dataset. The results of this research are significant to Local LLMs and AI R&D, especially sentiment analysis. With the advancements of AI programs such as ChatGPT and DeepSeek, getting an idea about the perception of the public toward these programs becomes crucial in guiding their development and use.

**7. Conclusion:**

This study contributes to the growing body of public opinion on LLM by offering an in-depth commentary of Reddit discussions on ChatGPT and DeepSeek. Through sentiment analysis and topic modeling, the study gives valuable insights on how the users perceive these technologies, what concerns they have about them, and what motivates their trust in AI. Public opinion needs to be understood in order to guide the ethical development of AI systems and their positive application in society. Continued research must continue examining users' attitudes and observing how their opinions evolve as LLMs become more part of everyday life.